\documentclass[a4paper,11pt]{article}

\usepackage{jinstpub} 
\usepackage{graphicx}
\usepackage{subfigure}
\usepackage{verbatim}
\usepackage{url}

\title{\boldmath A fast tunable driver of light source for the TRIDENT Pathfinder experiment}

\author[a,b]{J.N. Tang,}
\author[a,b,1,2]{W.H. Wu,\note{Corresponding author.}\note{These authors contributed equally.}}
\author[a,b,2]{L. Li,}
\author[c]{P. Miao,}
\author[a,b]{Z.Y. Sun,}
\author[a,b]{M.X. Wang,}
\author[d,b]{and D.L. Xu}

\affiliation[a]{Institute of Nuclear and Particle Physics, School of Physics and Astronomy, Shanghai Jiao Tong University,\\800 Dongchuan Road, Shanghai, China}
\affiliation[b]{Key Laboratory for Particle Astrophysics and Cosmology (MoE), Shanghai Key Laboratory for Particle Physics and Cosmology,\\800 Dongchuan Road, Shanghai, China}
\affiliation[c]{Department of Modern Physics, University of Science and Technology of China,\\96 Jinzhai Road, Hefei, China}
\affiliation[d]{Tsung-Dao Lee Institute, Shanghai Jiao Tong University,\\520 Shengrong Road, Shanghai, China}

\emailAdd{wuweihao@sjtu.edu.cn}

\abstract{TRIDENT (The tRopIcal DEep-sea Neutrino Telescope) is a proposed next-generation neutrino telescope to be constructed in the South China Sea. In September 2021, the TRIDENT Pathfinder experiment (TRIDENT EXplorer, T-REX for short) was conducted to evaluate the \textit{in-situ} optical properties of seawater. The T-REX experiment deployed three digital optical modules at a depth of 3420 meters, including a light emitter module (LEM) and two light receiver modules (LRMs) equipped with photomultiplier tubes (PMTs) and cameras to detect light signals. The LEM emits light in pulsing and steady modes. It features a fast tunable driver to activate light-emitting diodes (LEDs) that emit nanosecond-width light pulses with tunable intensity. The PMTs in the LRM receive single photo-electron (SPE) signals with an average photon number of approximately 0.3 per 1-microsecond time window, which is used to measure the arrival time distribution of the SPE signals. The fast tunable driver can be remotely controlled in real-time by the data acquisition system onboard the research vessel, allowing for convenient adjustments to the driver's parameters and facilitating the acquisition of high-quality experimental data. This paper describes the requirements, design scheme, and test results of the fast tunable driver, highlighting its successful implementation in the T-REX experiment and its potential for future deep-sea experiments.}

\keywords{Neutrino detectors; Analogue electronic circuits; Control and monitor systems online}

\arxivnumber{2305.01967} 

\proceeding{N$^{\text{th}}$ Workshop on X\\
  when\\
  where}

\begin{document}
\maketitle
\flushbottom

\section{Introduction}
\label{sec:intro} 

Neutrinos are extremely light and electrically neutral leptons that can only interact through gravity and weak force. Unlike other charged particles which can be deflected by electromagnetic fields, neutrinos can propagate through long distances without deflection and point back to their origin. Therefore, cosmic neutrinos are outstanding messengers of the extreme Universe and play a unique role in multi-messenger astronomy. Due to its extremely small interaction cross-section, in order to capture the high-energy astrophysical neutrinos, scientists must build neutrino telescopes with very large volumes of medium for neutrinos to interact within. These neutrinos pass through and interact with the detector medium, potentially producing high-energy secondary charged particles. If the velocity of the charged particle is faster than the velocity of light in this medium, Cherenkov radiation will be emitted. Using detector arrays with photo-multiplier tubes (PMTs) to detect the Cherenkov photons allows us to reconstruct the direction and energy of the primary cosmic neutrinos~\cite{aartsen2014}.

\begin{figure}[htbp]
	\centering 
	\includegraphics[width=.7\textwidth]{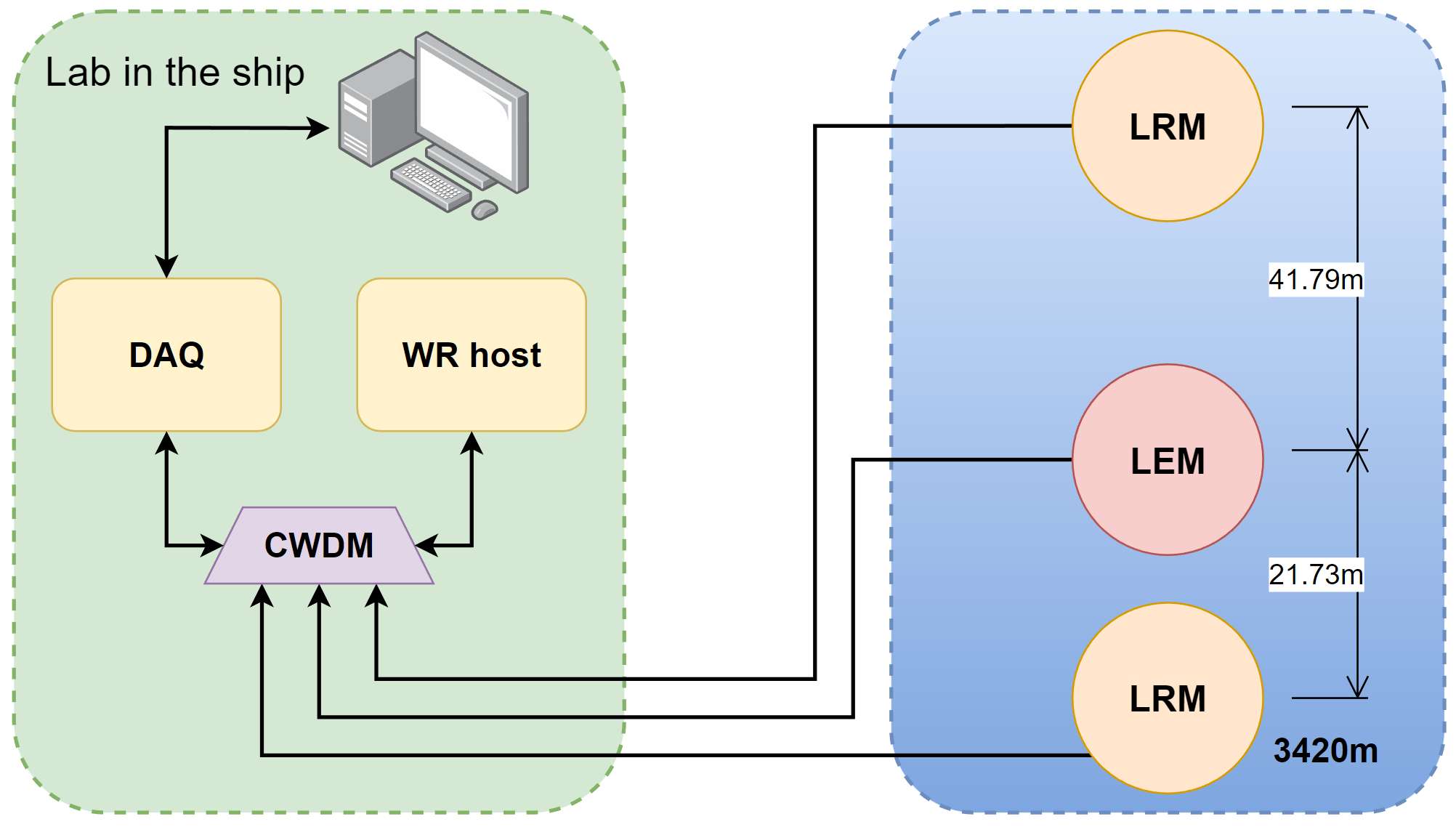}
	\caption{\label{fig:e_o_d}Block diagram of the T-REX electronics system. The part within the left dashed box is located in the laboratory on the back deck of the research vessel. The right part is placed 3420 meters underwater by an umbilical cable.}
\end{figure}

TRIDENT~\cite{TRIDENT_paper} (The tRopIcal DEep-sea Neutrino Telescope) is a next-generation neutrino telescope designed to be located at depths approximately between 2800 m and 3500 m in the northeast part of the South China Sea. The TRIDENT Pathfinder experiment (TRIDENT EXplorer, T-REX for short) aims to measure the \textit{in-situ} optical properties of seawater, the ambient background light, and the marine environment at the selected site. To achieve these objectives, the T-REX detector system comprises three digital optical modules that are connected by electro-optical cables in series, as illustrated in Figure~\ref{fig:e_o_d}. The White Rabbit (WR) protocol is used to synchronize clocks between the three digital optical modules~\cite{lipinski2011white}\cite{moreira2009white}. The middle light emitter module (LEM) emits pulsing and steady light from light-emitting diodes (LEDs) with various wavelengths. The PMTs and cameras in the upper and lower light receiver modules (LRMs) detect the pulsing and steady light, respectively.

The T-REX experiment utilizes a relative measurement strategy to decode the properties of seawater. To achieve this, the experiment is designed with upper and lower LRMs placed at unequal distances from the light source - specifically, at distances of 41.79$\pm$0.04 meters and 21.73$\pm$0.02 meters~\cite{TRIDENT_paper}, respectively - as shown in Figure~\ref{fig:e_o_d}. These distances are comparable to the typical attenuation length ($\sim$ 20 m) of seawater. The PMTs in the LRM detect the pulsed light from the LEM. The position of the received PMT signals in the event window indicates the arrival time of the photons, since the LRM and LEM are synchronized. The pulsing light emitted from LEM can be scattered by the seawater, resulting in the expected arrival time of photons reaching the PMT spreading out over a certain width. The light scattering properties of seawater can be obtained by measuring and comparing the time width between the two LRMs. The camera system is composed of a 5-megapixel monochrome camera and a Raspberry-4Pi module, which detects the steady light from the LEM. The light absorption properties can be determined by comparing the light intensity detected by the upper and lower cameras. 

This measurement strategy presents some challenges for the middle LEM. The actual attenuation conditions in the deep sea are unknown, requiring a light pulser with adjustable intensity to adapt flexibly to different scenarios. Additionally, to achieve accurate measurements of the arrival time distribution of emitted photons, the width of the light pulse needs to be on the order of a few nanoseconds, which is significantly smaller than the width of the distribution. To address these challenges, a fast and tunable light source driver has been designed to meet the requirements of T-REX. 

The detailed requirements of the fast tunable driver are described in section~\ref{sec:requirement}. The design scheme and the performance of the fast tunable driver are introduced in section~\ref{sec:design_scheme} and section~\ref{sec:results}. Finally, the conclusions are presented in section~\ref{sec:conclusions}.


\section{Performance Requirements}
\label{sec:requirement} 

The fast tunable driver consists of two kinds of LED drivers: the pulsing LED driver for the PMT detection system and the steady LED driver for the camera detection system. The pulsing LED driver should satisfy the following requirements:

\begin{itemize}
	\item[1.]The width of LED pulses and the transit time spread (TTS) of photomultiplier tubes (PMTs) can both contribute to broadening the distribution of photon arrival times. To minimize the impact on the arrival time distribution, the pulse width of the LED must be significantly narrower than the width of the arrival time distribution (on the order of 100 ns). Therefore, the time width of the LED pulses should be on the order of nanoseconds.
	\item[2.]The light intensity of a single pulse should be adjustable in a wide range. In the T-REX experiments, the arrival time distribution of PMT's single photo-electron (SPE) signals is used to determine the light scattering properties. The proportion of SPE signals is controlled by adjusting the light pulse intensity. However, the suitable light intensity is related to many factors, such as the sensitive area of PMT, the properties of seawater, etc. Thus, it is difficult to accurately determine the light intensity suitable for the deep-sea experiment. Variations in the deep-ocean environment further complicate the situation.
	\item[3.]To simplify the process of adjusting the light intensity and save time in the deep-sea experiment, an automatic bias voltage scan has been developed to rapidly determine the appropriate light intensity of the pulsing LEDs.
	\item[4.]The pulsing LED driver should be controlled in real-time by the T-REX host computer on the research vessel, allowing for the control of the wavelength, light intensity, and pulse frequency of the LEDs.
	\item[5.]The three wavelengths of LEDs (405 nm, 450 nm, and 525 nm) are selected to emit pulsing light. These wavelengths are representative of the Cherenkov radiation spectrum in the seawater.
\end{itemize}

The steady LED driver should correspond to the following requirements:
\begin{itemize}
    \item[1.]The steady LEDs are driven in the constant current mode. The temperature of an LED increases as it starts operating, causing a leftward shift in its voltage-current (V-I) characteristic curve. If the driving voltage of the LED is kept constant, the current flowing through the LED increases as the temperature rises, resulting in an increase in the brightness of the LED. The camera detection system requires the LED brightness to be constant over time.
    \item[2.]Fifteen steady LEDs with three wavelengths (405 nm, 460 nm, and 525 nm) are selected to emit steady light. Each five LEDs with the same wavelength are driven in series to ensure the same brightness of each LED.
\end{itemize}


\section{Design scheme}
\label{sec:design_scheme}

\begin{figure}[htbp]
	\centering
	\includegraphics[width=.95\textwidth]{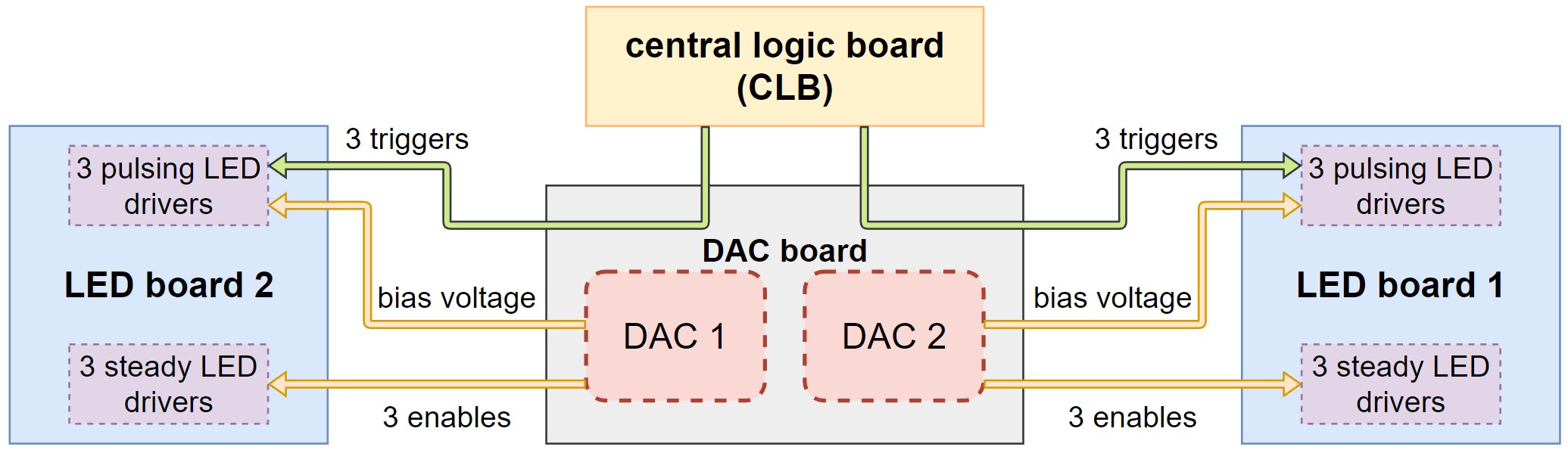}
	\caption{\label{fig:Block_diagram_FastLightSource}Block diagram of the fast tunable driver of light source in the LEM.}
\end{figure}

The fast tunable driver, as shown in Figure~\ref{fig:Block_diagram_FastLightSource}, comprises two LED boards, a central logic board (CLB), and a DAC (Digital to Analog Converter) board. The LED boards are symmetrically placed in the two hemispheres. The CLB receives and decodes the commands from the host computer to control the working status of the fast tunable driver, like lighting mode switching, pulsing light intensity adjustment.

\begin{figure}[htbp]
	\centering
	\includegraphics[height=.5\textwidth]{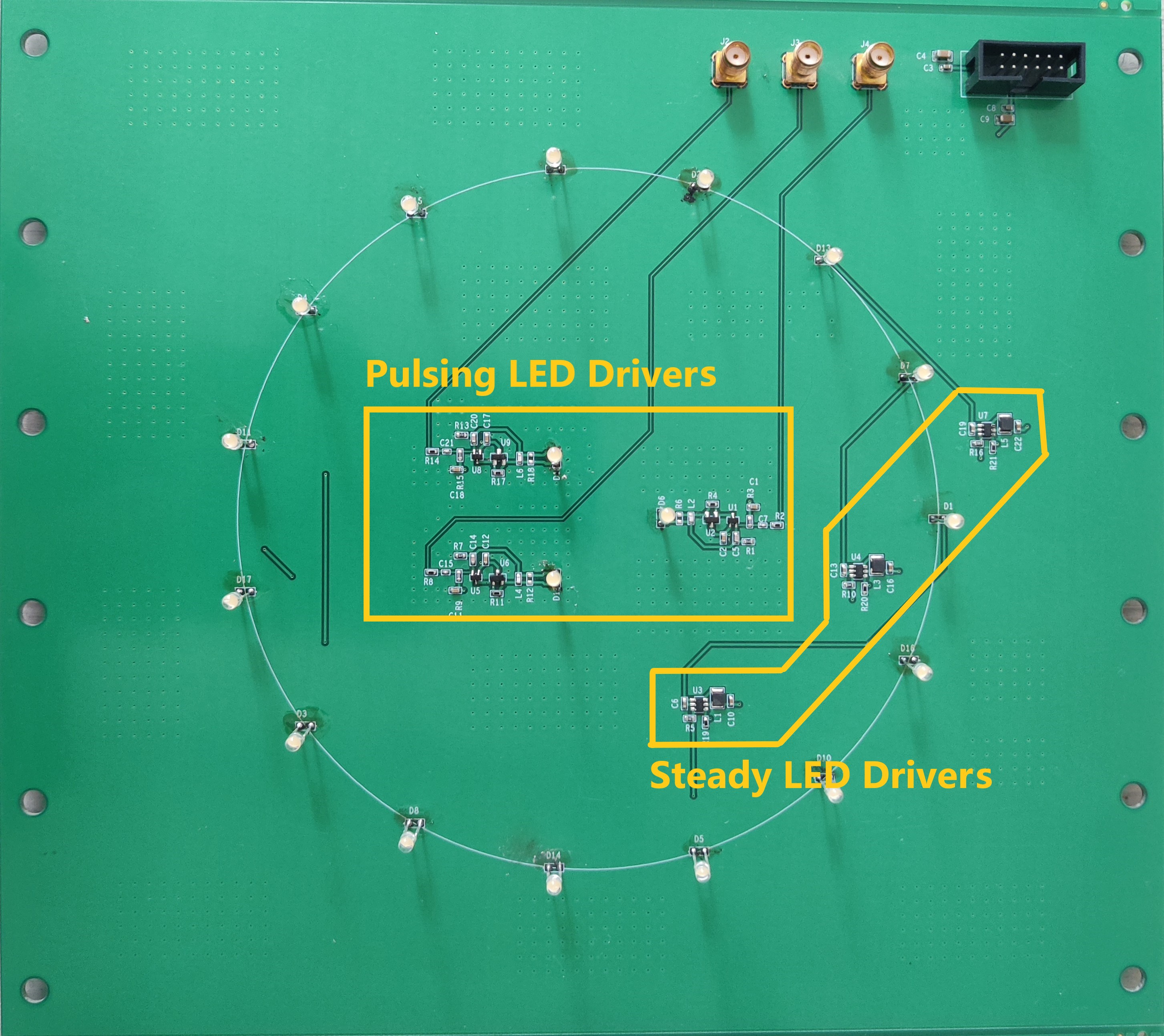}
	\caption{\label{fig:LED_layout}A photograph of the LED board with markings indicating the positions of the three pulsing LED drivers and three steady LED drivers.}
\end{figure}

Figure~\ref{fig:LED_layout} shows the LED board employed in the fast tunable driver. The middle part of the LED board features three pulsing LED drivers, each of which drives a pulsing LED. Additionally, there are three steady LED drivers, which operate five steady LEDs each. These steady LEDs are evenly distributed on a circular plane, with identical angular spacings. More details on the design principle of the LED system is presented in~\cite{li2023light}.

\subsection{Nanosecond-width pulsing LED driver}
\label{subsec:pulsed_led}

\begin{figure}[htbp]
	\centering
  	\includegraphics[width=.65\textwidth]{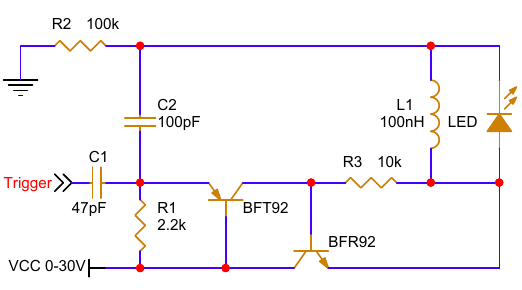}
    \caption{\label{fig:pulser_sch}The schematics of the nanosecond-width LED pulser. The complementary transistors pair determines the opening speed of the switch. With a certain bias voltage, The C2 and L1 determine the light pulse width and intensity.}
\end{figure}

The pulsing LED driver employs the capacitor discharging method to generate narrow pulses through a switch. A complementary pair regenerative switch~\cite{veledar2005}, originally proposed by J.~S.~Kapustinsky~\cite{kapustinsky1985}, is used in our design to generate nanosecond pulses. Due to its advantages of high performance, simplicity, and convenience, it is widely used in calibrations of high energy neutrino telescopes, such as IceCube~\cite{jurkovivc2016} in Antarctica, KM3NeT~\cite{real2019nanobeacon} in the Mediterranean Sea, and the Baikal-GVD~\cite{avrorin2019neutrino} in Lake Baikal.

\begin{figure}[htbp]
	\centering
	\includegraphics[width=.75\textwidth]{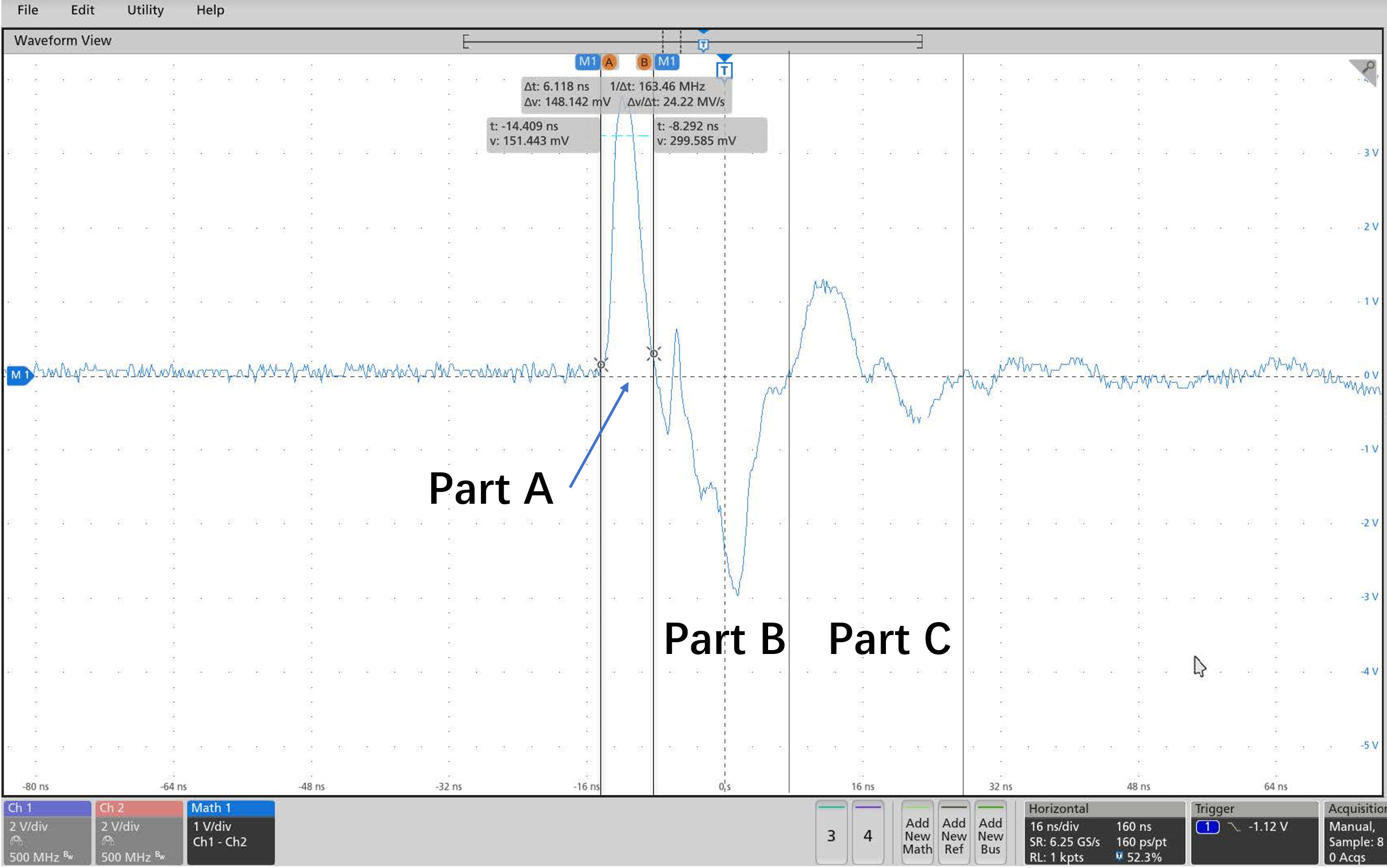}
	\caption{\label{fig:LED_pulse}The pulse shape of the nanosecond-width pulsing LED driver tested on an oscilloscope.}
\end{figure}

The schematic of the Kapustinsky pulser is shown in Figure~\ref{fig:pulser_sch}. The FPGA in the CLB generates a trigger signal, which is a 50\% duty square wave with an amplitude of 3.3 V. When the rising edge of the trigger signal passes through capacitor C1 and reaches the emitter of the PNP transistor (BFT92), the emitter voltage rises, causing an increase in the base current of the NPN transistor (BFR92). The further increase in base current of the NPN transistor increases the base current of the PNP transistor. This positive feedback process quickly opens the switch composed of the two transistors. The switch turning on provides a low impedance path for C2 to discharge through the LED and create a current pulse. An inductor (L1) connected in parallel with the LED is used to reduce the pulse width~\cite{lubsandorzhiev2006}. Figure~\ref{fig:LED_pulse} shows a voltage pulse across the LED tested by an oscilloscope. The pulse can be separated into three parts. Part A corresponds to the generation of the light pulse by the LED, and its typical width is about 3 ns. Parts B and C are formed at the resonance of L1 and C2. Part B is opposite to the forward voltage of the LED, while part C has a small amplitude, neither of which result in light emission.

\begin{figure}[htbp]
	\centering
	\subfigure[Pulse shapes with five different bias voltages.]{
  	    \includegraphics[width=7cm,height=7cm]{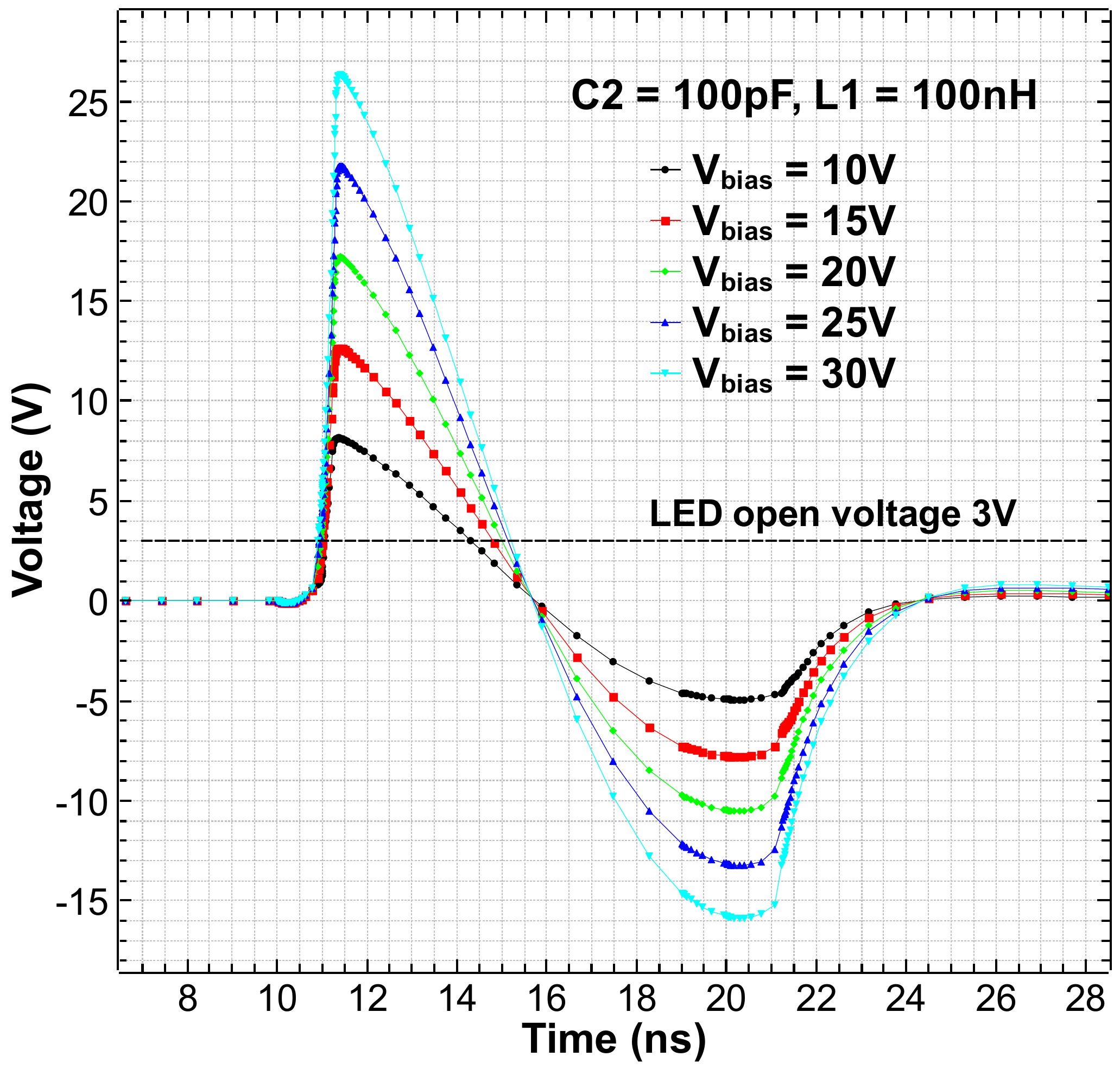}
   	    \label{fig:LED_1}
    }
    \subfigure[Light intensity with respect to bias voltages.]{
  	    \includegraphics[width=7cm,height=7cm]{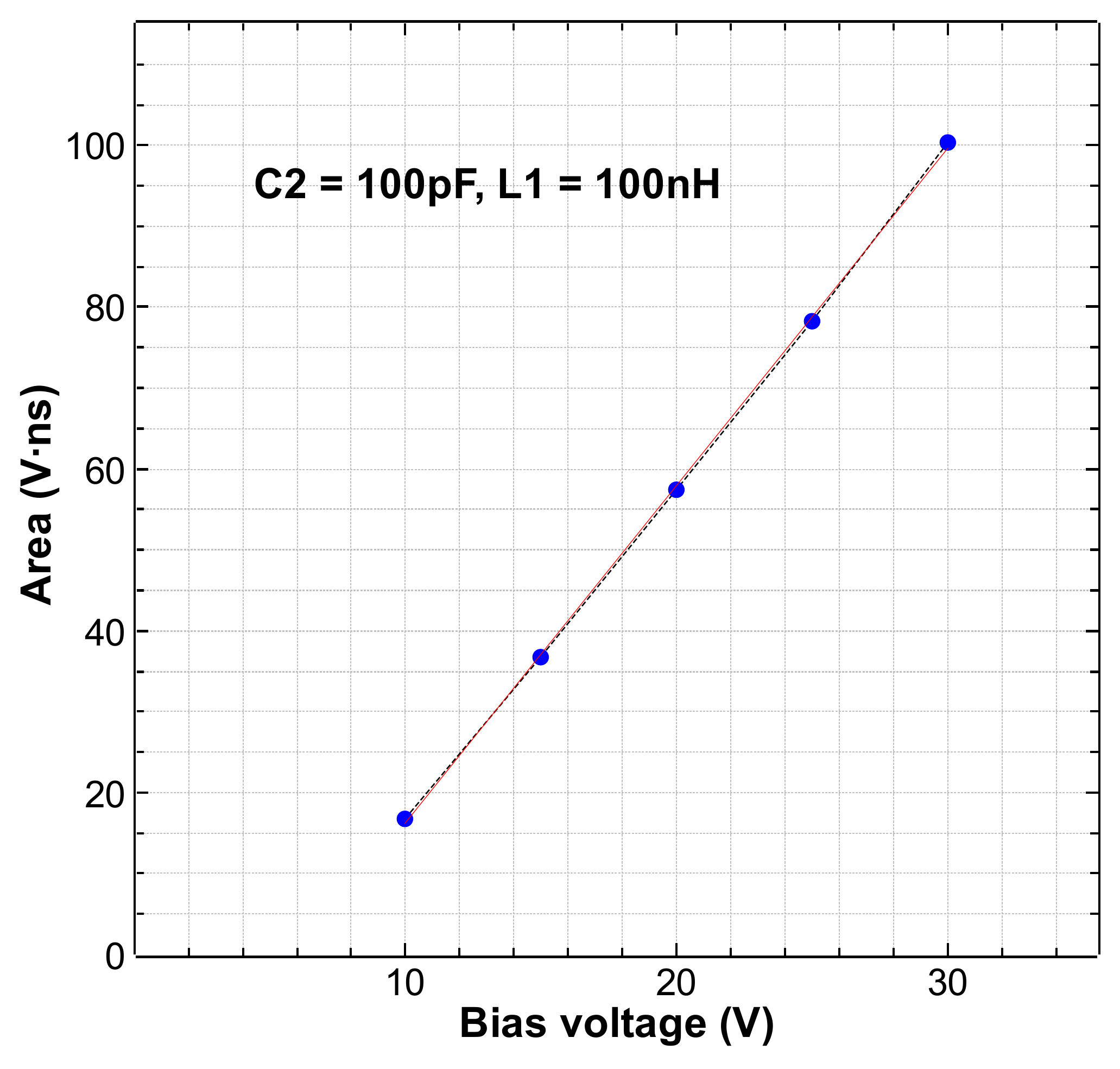}
   	    \label{fig:LED_2}
    }
	\subfigure[Pulse shapes with five different inductor L1.]{
		\includegraphics[width=7cm,height=7cm]{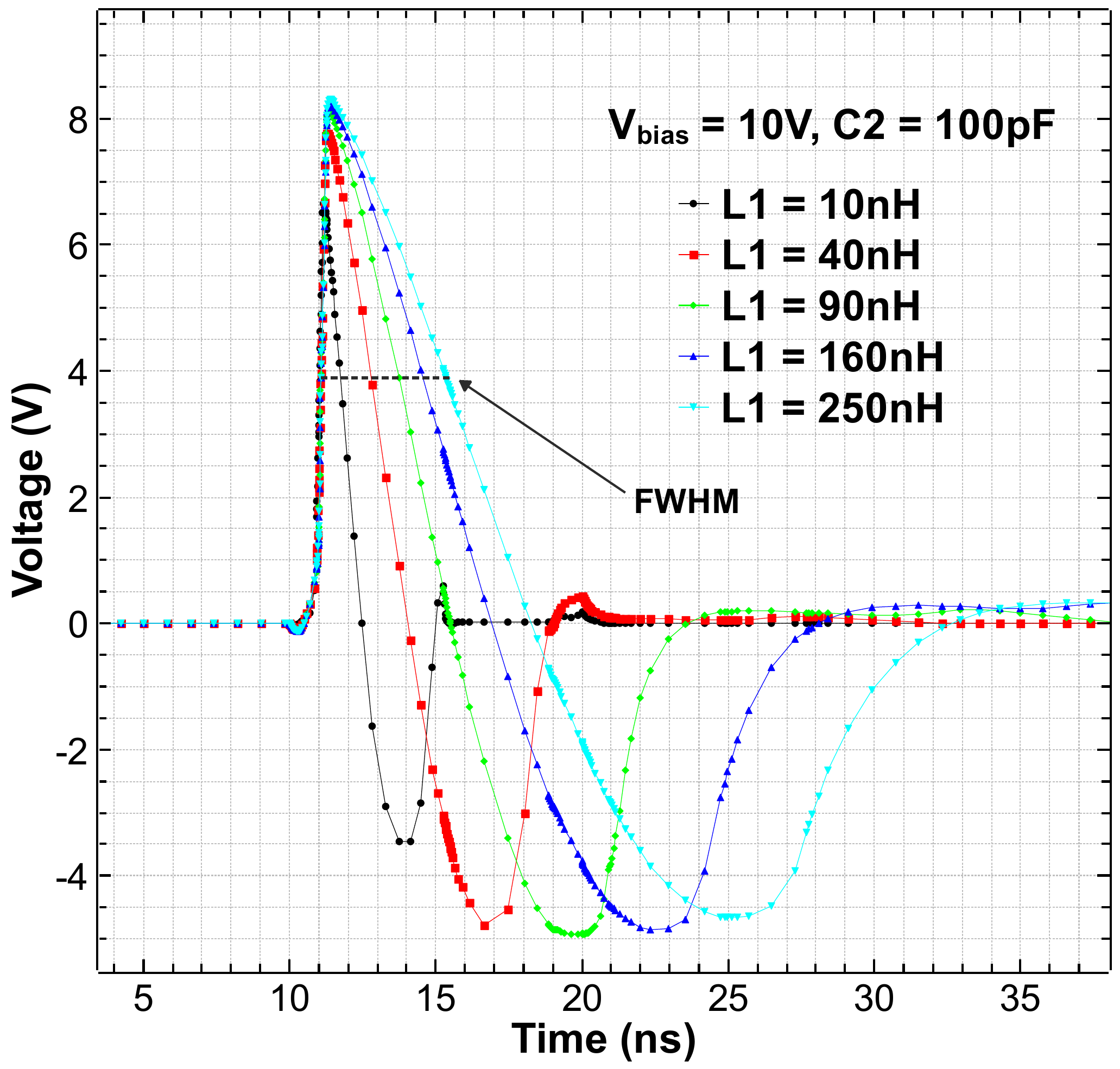}
		\label{fig:LED_3}
	}
	\subfigure[Pulse FWHM with respect to $\sqrt{L_{1}C_{2}}$.]{
		\includegraphics[width=7cm,height=7cm]{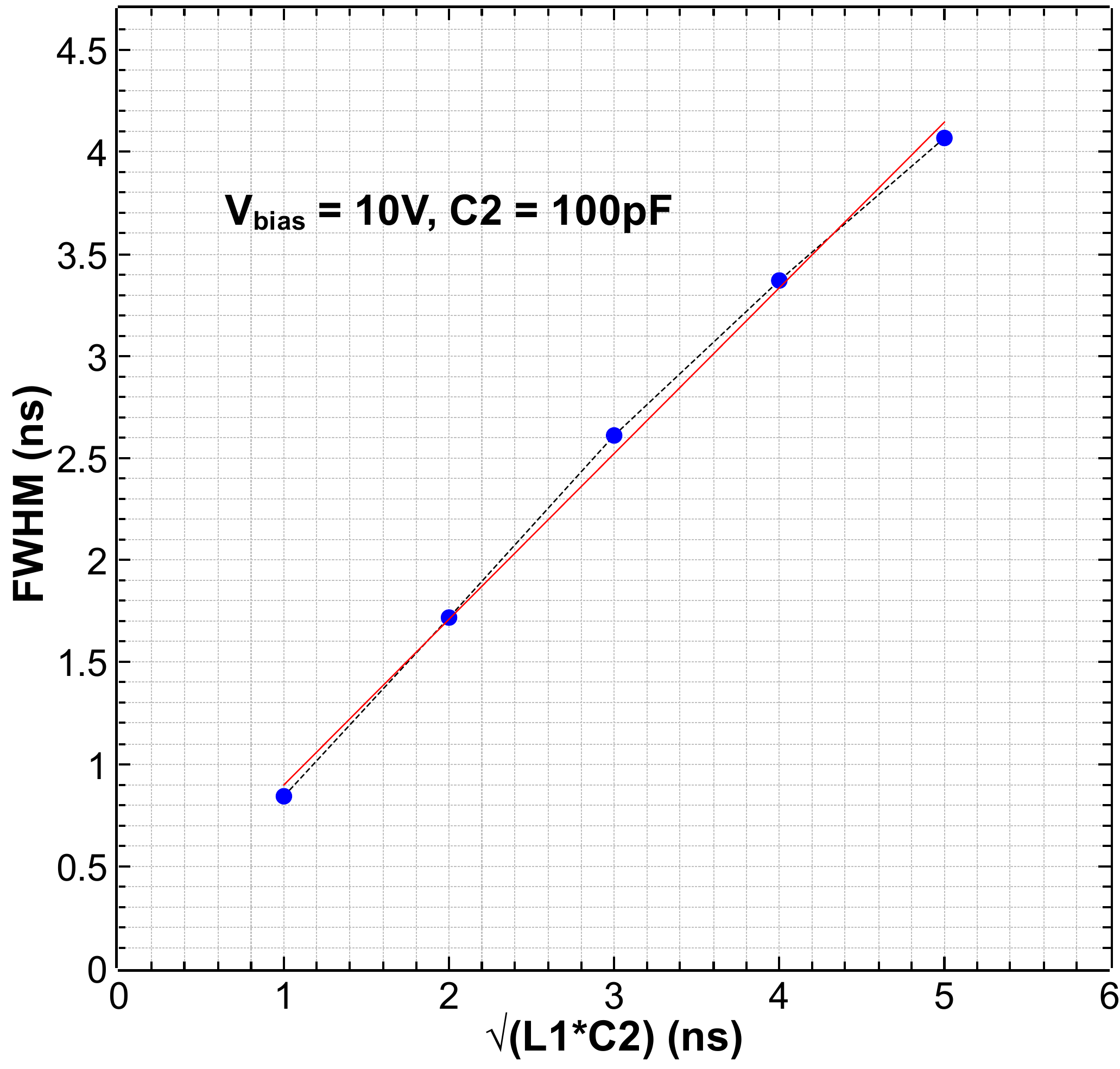}
		\label{fig:LED_4}
	}
    \caption{\label{fig:LED_L&R}The pulse shapes generated by the nanosecond-width pulsing LED driver.}
\end{figure}

The width of the light pulse, generated by the pulsing LED driver, is dependent on the charge capacitor C2 and the inductor L1. The external bias voltage is responsible for charging the storage capacitor C2 and determining the light intensity. The performance of the pulsing LED driver has been simulated in the PSpice and the results are displayed in Figure~\ref{fig:LED_L&R}. The pulse width is defined as the full width at half maximum (FWHM) of the waveform, while the pulse intensity is given by the integral area enclosed by the waveform and the baseline of the LED open voltage. Figure~\ref{fig:LED_1} demonstrates that the bias voltage does not significantly impact the pulse width. Figure~\ref{fig:LED_2} illustrates the linear relationship between the light intensity and the bias voltage. The relationship between the pulse width and L1 is presented in Figures~\ref{fig:LED_3} and ~\ref{fig:LED_4}, where the pulse width is linearly scaled with respect to $\sqrt{L1C2}$. In the T-REX experiment, C2 and L1 are set to 100pF and 100nH respectively to achieve sufficient pulse intensity and a pulse width of $\sim$ 3 ns.

\begin{figure}[htbp]
	\centering
	\includegraphics[width=.5\textwidth]{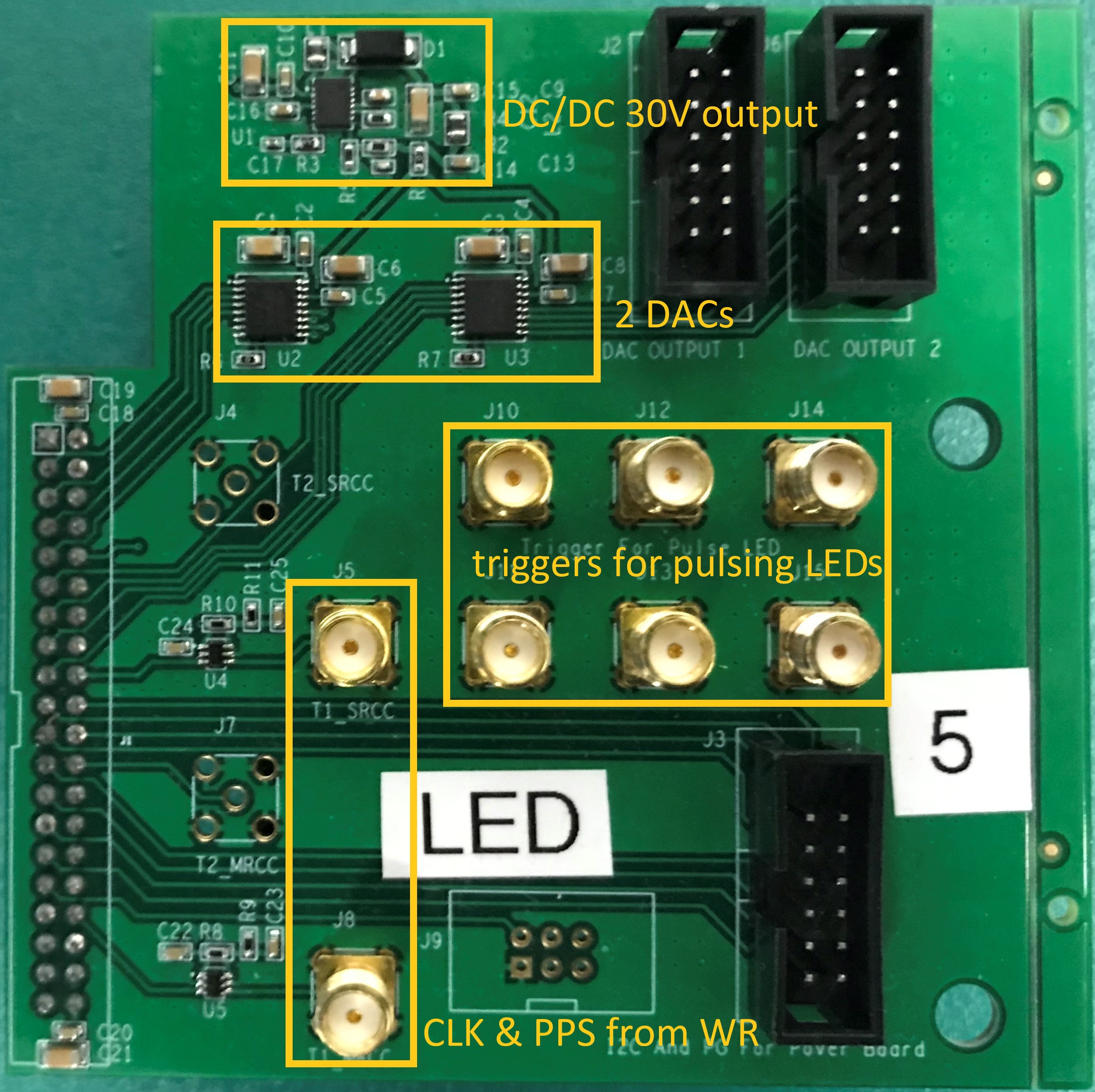}
	\caption{\label{fig:DAC_layout}A photograph of the DAC mezzanine board reveals the presence of two DACs and one DC/DC converter. There are six SMA connectors that provide triggers to the LED boards.}
\end{figure}

The light intensity of the pulsing LED can be adjusted by the external bias voltage. To achieve this, a DAC mezzanine board has been designed to provide the necessary bias voltage for the nanosecond LED pulser and enable switching of light modes. Figure~\ref{fig:DAC_layout} illustrates the photo of the DAC board. The key components of the DAC board include two 12-bit, four-channel high voltage DAC chips (AD5504~\cite{AD5504DAC} from Analog Device). These DAC chips can be configured via a serial peripheral interface (SPI) with the FPGA in the CLB. Additionally, a DC/DC converter is utilized to boost the voltage from 5 V to 30 V, which is adopted to provide a reference voltage to the DAC chip. The DAC output range is 0 V to 30 V, with a regulation step size of 7 mV, allowing for a large intensity range and fine adjustment of the light pulse intensity.

\subsection{Steady LED driver}
\label{subsec:const_led}

The LED board employs three steady LED drivers that collectively drive fifteen LEDs at three different wavelengths (405 nm, 460 nm, and 525 nm). The LEDs are arranged in a circle with a radius of 7.5 cm and equal angular spacing. Each set of five LEDs with the same wavelength is driven by one steady LED driver. To ensure consistent brightness, a constant current LED driver LT3465 from Analog Devices, is employed. The LT3465 is a step-up DC/DC converter capable of driving up to five LEDs in series.

\subsection{Automatic voltage scan in the software}
\label{subsec:voltage_scan}

\begin{figure}[htbp]
	\centering
	\includegraphics[width=.9\textwidth]{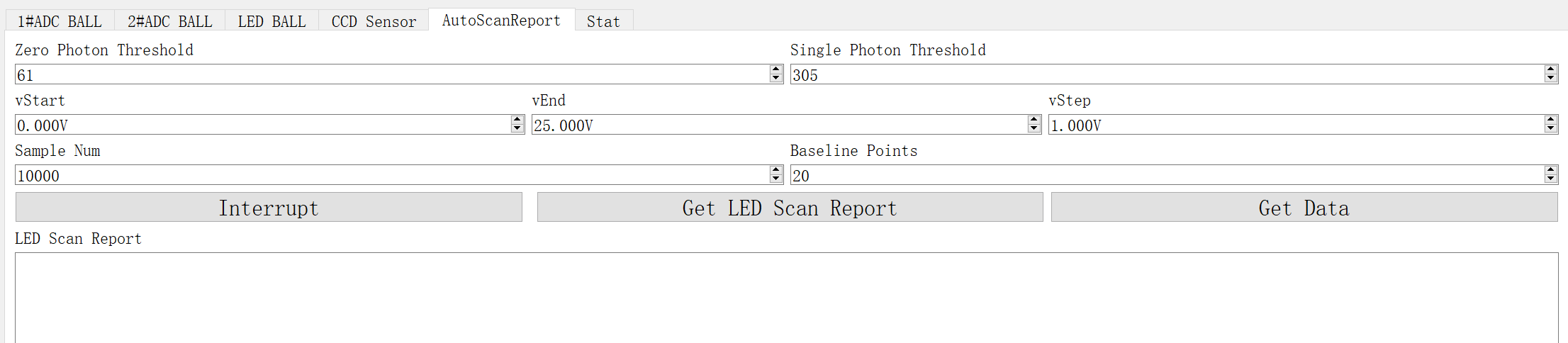}
	\caption{\label{fig:AutoScanUI}A picture of the voltage scan user interface.}
\end{figure}

In the deep-sea experiment, it is important to set an appropriate proportion of the SPE signals in the measurement for scattering properties of the seawater. The method of manually adjusting the bias voltage to change the light intensity is time-consuming. The power limitations of the battery in the deep-sea experiment further restrict the available experimental time. Thus, an automatic bias voltage scan for the pulsing LED driver is developed in the software of the host computer to quickly determine the optimal bias voltage. The user interface (UI) of the software is shown in Figure~\ref{fig:AutoScanUI}. First, the start and end points of the scanning voltage as well as the step size are specified. At a scanned voltage, the signals from the PMTs with amplitudes falling between the noise threshold and the multi-photon signal threshold are treated as SPE signals. Subsequently, the proportion of SPE signals corresponding to the scanning voltage is calculated and can be viewed in the 'LED Scan Report' section of the user interface (UI). When utilizing the LED voltage scanning feature, it is advisable to initiate with an extensive scanning range and substantial step size to initially approximate the required voltage. Following this, the step size should be reduced to fine-tune and identify the precise bias voltage. Employing this approach can optimize the time efficiency and facilitate precise bias voltage determination.



\section{Test results}
\label{sec:results}

\subsection{Laboratory commissioning results}
\label{sec:test_system}

\begin{figure}[htbp]
	\centering
	\subfigure[]{
  	    \includegraphics[width=.45\textwidth]{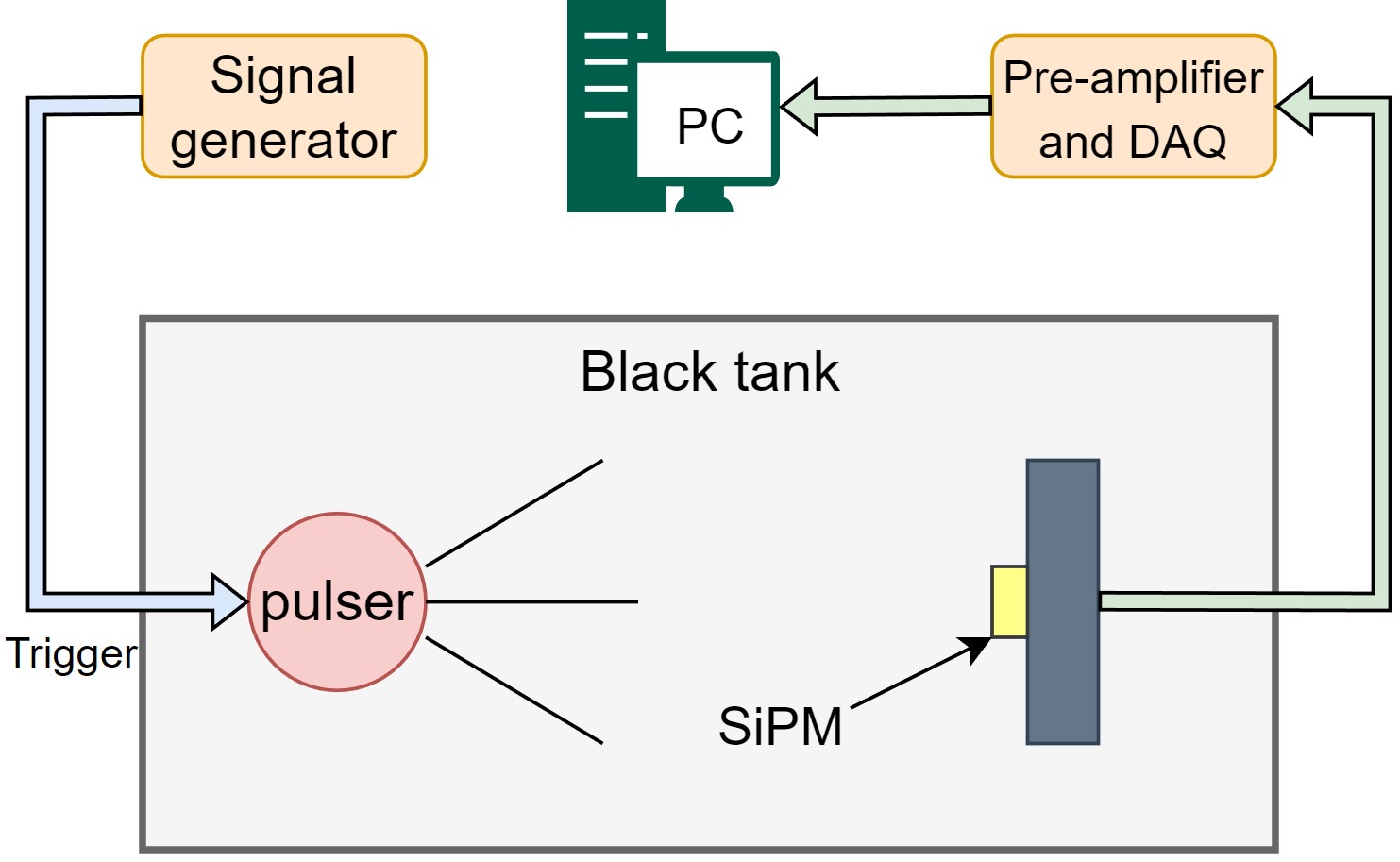}
   	    \label{fig:SiPMtest_block}
    }
    \subfigure[]{
  	    \includegraphics[width=.51\textwidth]{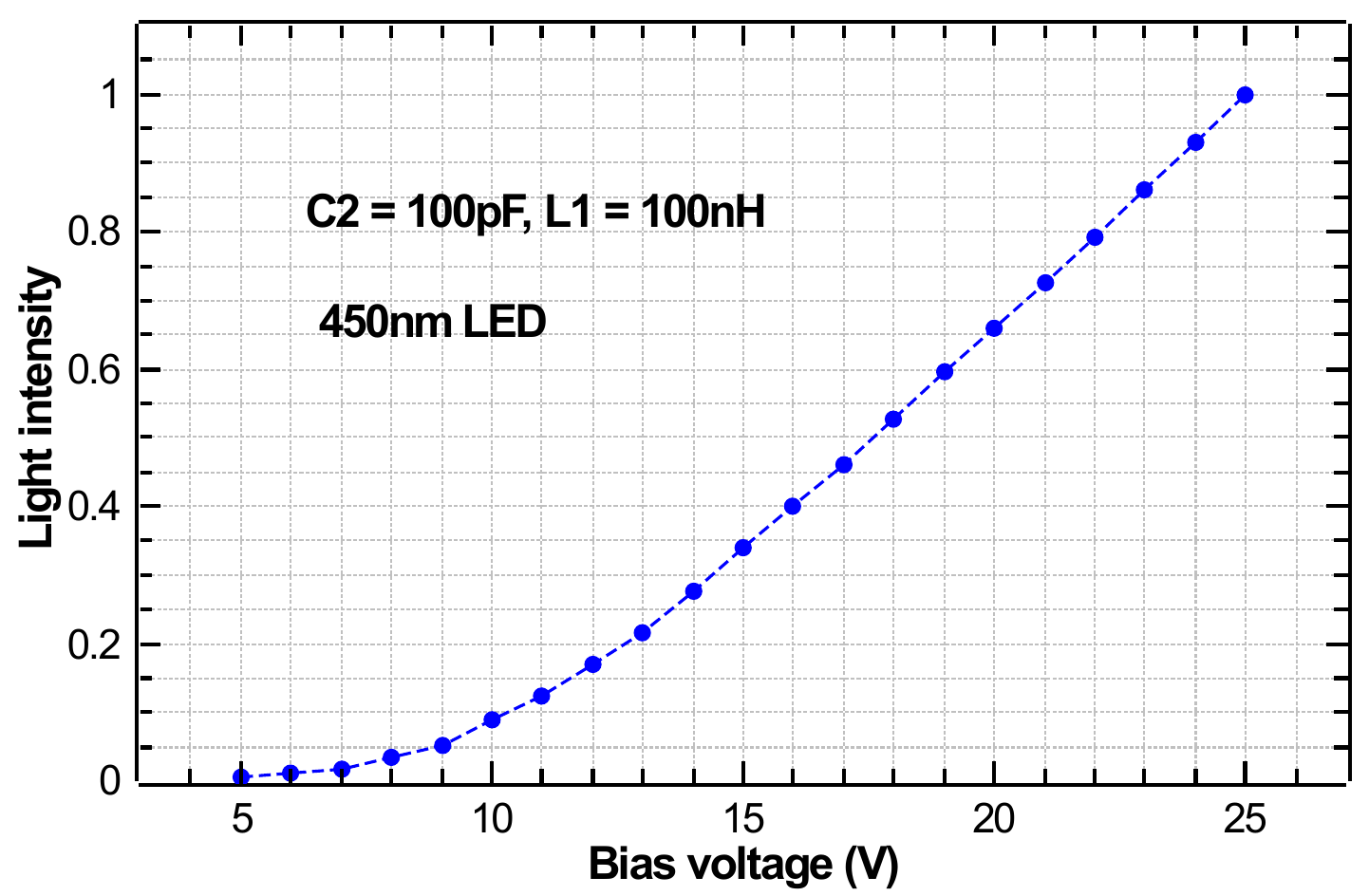}
   	    \label{fig:LightIntensity_VS_Volt}
    }
    \caption{\label{fig:sipm_test}Figure (a) shows the block diagram of the light intensity test system, while Figure (b) illustrates the relative light intensity of a pulsing LED driver as a function of bias voltage. The light intensity is normalized based on the light intensity measured at the bias voltage of 25V.}
\end{figure}

A laboratory test is carried out to validate the performance of the fast tunable driver before the deep-sea experiment. A 450 nm LED and a SiPM (silicon photomultiplier) S13360-1325CS~\cite{1325CS} from Hamamatsu are employed to measure the relative light intensity with different bias voltages~\cite{henningsen2020self}, as presented in Figure~\ref{fig:SiPMtest_block}. A capacitor C2 of 100 pF and an inductor L1 of 100 nH are implemented in the setup, and the pulse width is around 3 ns. The light intensity and bias voltage exhibit a favorable linear relationship in the range of 10 V to 30 V, as illustrated in Figure~\ref{fig:LightIntensity_VS_Volt}. This result indicates that the light pulse intensity can be adjusted in a wide range to satisfy the design requirements.

\begin{figure}[htbp]
	\centering
	\includegraphics[width=.7\textwidth]{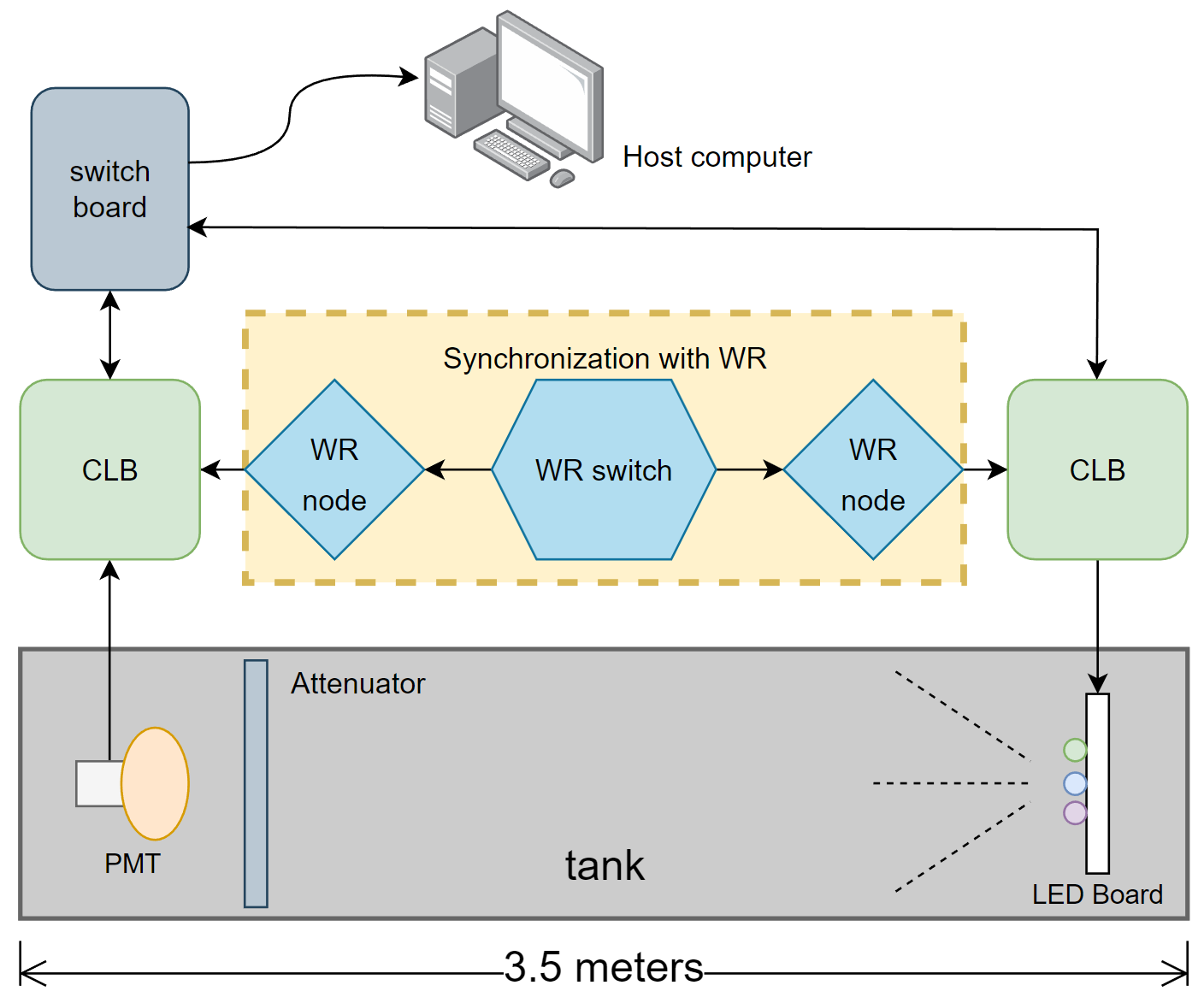}
	\caption{\label{fig:block_testsys}Block diagram of the driver test system. The attenuator is placed close to the PMT.}
\end{figure}

The pulsing LED driver is also tested with the PMTs used in the deep-sea experiment. The experimental setup, as shown in Figure~\ref{fig:block_testsys}, consists of a LED board and a PMT placed on opposite ends of a tank. The tank is covered with blackout cloth to exclude external ambient light interference. A CLB is utilized as the control and readout system, and a WR system is employed for time synchronization.

The distance between the LEM and the upper LRM in the deep-sea is 41.79 meters. Limited by the test condition, and the distance between the LED board and the PMT in the laboratory test is only about 2 meter. The attenuation factors of the light intensity in the laboratory and the actual experiment differ by a factor of approximately 400. Considering the absorption of light by seawater, a light attenuator with an attenuation factor of 500 is placed between the LED board and the PMT, closer to the PMT.

The photon number within each DAQ time window should follow a Poisson distribution, which is expressed as:
\begin{equation*}
P(k)=\frac{e^{-\lambda} \lambda^{k}}{k!}
\end{equation*}
Here, $k$ denotes the number of detected photons, and $\lambda$ represents the mean value of detected photons within the time window. In the laboratory test, the average photon number ($\lambda$) is set to 0.1. Out of all the signals, the proportion of single-photon events (SPE signals) is around 9\%, and the proportion of multi-photon events is approximately 1\%. The remaining 90\% of the signals exhibit no photo-electron activity. The value of $\lambda$ is selected to ensure an adequate number of SPE signals while keeping the occurrence of multi-photon events low. During testing, the proportion of SPE signals received by the PMTs can be controlled by adjusting the bias voltage of the pulsing LED driver.

\begin{figure}[htbp]
	\centering
	\includegraphics[width=.4\textwidth]{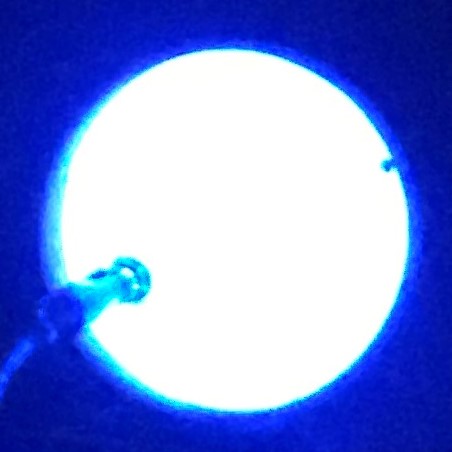}
	\caption{\label{fig:LEM_photo}A photograph of the LEM emitting steady light.}
\end{figure}

The performance of the steady LED driver is also evaluated in the laboratory. To fulfill the luminosity requirement of the camera system, a current of 20 mA is applied to the steady LEDs. The sphere's total anisotropy, quantified as 19.2\%, has been experimentally evaluated in the laboratory. The steady LEDs fulfill the requirements for isotropic and time-invariant light emission on the sphere. Figure~\ref{fig:LEM_photo} shows an image of the LEM working in steady mode.

\subsection{The T-REX results}
\label{sec:exp_results}

The T-REX, conducted in September 2021, deployed experimental apparatus at a depth of 3420 meters, successfully collected $\sim$ 1TB \textit{in-situ} optical experiment data in both the pulsing and steady modes. The fast tunable driver was instrumented in the light source module to initiate photon signals to PMTs located in the LRMs, enabling us to measure the arrival time distribution of the SPE signals.

We utilized an automated bias voltage scan to ascertain the light pulse intensity and the proportion of SPE signals from the pulsing LEDs. To obtain sufficient experiment data while considering the limitations of the DAQ bandwidth, the pulsing frequency is set to 10 kHz. The average photon count per event was fixed at 0.3, a value chosen to enhance the contribution of SPE signals while minimizing the occurrence of multi-photon events. The proposed distributions of the SPE signals' arrival time are discussed in detail in the scientific paper of the T-REX~\cite{TRIDENT_paper}, which were used to analyze the scattering properties of the deep-sea water.

Furthermore, the fast tunable driver operated in the steady mode and was used for the camera detection system to analyze the light absorption properties of seawater. The fast tunable driver exhibited stable performance and complied with the design requirements during the deep-sea experiment.


\section{Conclusions}
\label{sec:conclusions}

TRIDENT is a proposed next-generation neutrino telescope to be constructed in the South China Sea. Its pathfinder experiment, T-REX, has been conducted to evaluate the \textit{in-situ} optical properties of seawater. The fast tunable driver consisting of the pulsing and steady LED drivers is designed for the detection systems of T-REX. The pulsing LED driver provides nanosecond-width light pulses to the PMT system, whereas the steady LED driver provides constant brightness light to the camera system. The fast tunable driver has been successfully utilized in the T-REX experiment, demonstrating robustness and reliability. Its tunable intensity and nanosecond time width make it suitable for use in a variety of light detection experiments.

\acknowledgments
We thank Jun Guo for valuable comments and insightful discussions about the paper. This work was sponsored by Shanghai Pujiang Program (No.20PJ1409300), Shanghai Pilot Program for Basic Research – Shanghai Jiao Tong University (21TQ1400218), Yangyang Development Fund and the Oceanic Interdisciplinary Program of Shanghai Jiao Tong University (project number SL2022MS020).








\bibliographystyle{JHEP}
\bibliography{myref}
\end{document}